\documentclass[twocolumn,showpacs,preprintnumbers,amsmath,amssymb,superscriptaddress]{revtex4}
\usepackage{calc}
\usepackage{float}
\usepackage{graphicx}% Include figure files
\usepackage{bm}% bold math
\begin{document}

\title{Geometric friction directs cell migration}

\author{M. Le Berre}
\affiliation{Institut Curie, CNRS UMR 144, 26 rue d'Ulm, 75005, Paris, FRANCE} 
\affiliation{Equal contribution}

\author{Yan-Jun Liu}
\affiliation{Institut Curie, CNRS UMR 144, 26 rue d'Ulm, 75005, Paris, FRANCE} 
\affiliation{Equal contribution}

\author{ J. Hu}
\affiliation{Ecole Normale Supérieure, Department of Chemistry, UMR 8640, 24 rue Lhomond, 75005 Paris, FRANCE} 

\author{Paolo Maiuri}
\affiliation{Institut Curie, CNRS UMR 144, 26 rue d'Ulm, 75005, Paris, FRANCE} 

\author{O. B\'enichou}
\affiliation{Laboratoire de Physique Th\'eorique de la Mati\`ere Condens\'ee, UMR 7600 CNRS /UPMC, 4 Place Jussieu, 75255
Paris Cedex}

\author{R. Voituriez}
\affiliation{Laboratoire de Physique Th\'eorique de la Mati\`ere Condens\'ee, UMR 7600 CNRS /UPMC, 4 Place Jussieu, 75255
Paris Cedex}
\affiliation{Laboratoire Jean Perrin, FRE 3231 CNRS /UPMC, 4 Place Jussieu, 75255
Paris Cedex}
\affiliation{Corresponding authors}

\author{ Y. Chen}
\affiliation{Ecole Normale Supérieure, Department of Chemistry, UMR 8640, 24 rue Lhomond, 75005 Paris, FRANCE} 
\affiliation{Corresponding authors}

\author{M. Piel}
\affiliation{Institut Curie, CNRS UMR 144, 26 rue d'Ulm, 75005, Paris, FRANCE} 
\affiliation{Corresponding authors}

%\date{\today}

\begin{abstract}
In the absence of environmental cues, a migrating cell performs an isotropic random motion.  Recently, the breaking of this isotropy has been  observed when cells move in the presence of asymmetric adhesive  patterns. However, up to now the mechanisms at work to direct cell migration in such  environments remain unknown.  Here, we show that a non-adhesive surface with asymmetric micro-geometry consisting of dense arrays of tilted micro-pillars 
 can direct cell motion.  Our analysis reveals that most features of cell trajectories, including the bias,  can be reproduced by a simple  model of active Brownian particle in a ratchet potential, which we suggest originates from  a generic elastic interaction of the cell body with the environment. The observed guiding effect, independent of adhesion, is therefore robust and could be used  to direct cell migration both in vitro and in vivo.

 \end{abstract}

%\pacs{05.40.Jc, 05.40.Fb}

\maketitle
Cells in vivo migrate along specified directions to contribute to tissue development and homeostasis or to achieve specialized functions. This directional migration is essential for a large set of biological processes, including morphogenesis \cite{Ridley:2003fk},  wound healing \cite{Poujade:2007uq}, tumor spreading \cite{Huber:2005kx} or immune responses \cite{Faure-Andre:2008,Hawkins2009,Schumann:2010vn}.  Among the numerous examples of taxis,  it is well established that asymmetries in the mechanical properties of the environment, such as stiffness  \cite{Saez:2007ly,Lo:2000ve}  or adhesion gradients \cite{Kim:2009qf}, can direct cell migration. Recently, it has been shown that  asymmetric adhesive micro-patterns or asymmetric geometrical obstacles can bias the direction of cell movement \cite{Mahmud:2009bh,Jiang:2005dq,Ohnuma:2009cr}. From the physical point of view, a cell is self-propelled and should therefore be viewed as a particle out of thermal equilibrium. As such, it is expected on general grounds \cite{Julicher:1997fk,Reimann:2002ve,Angelani:2011fk} that in an asymmetric environment its long term motion should be biased, according to a ratchet and pawl mechanism. However, up to now the microscopic processes at work to direct cell migration in asymmetric environments remain elusive, so that even the direction of migration often stays unpredictable in available set-ups.

Here, we propose to assay the capacity of  cells to be directed by purely geometrical factors in their environment, using micro-structured asymmetric substrates on which they cannot adhere. 
\begin{figure}
\includegraphics[width=9cm]{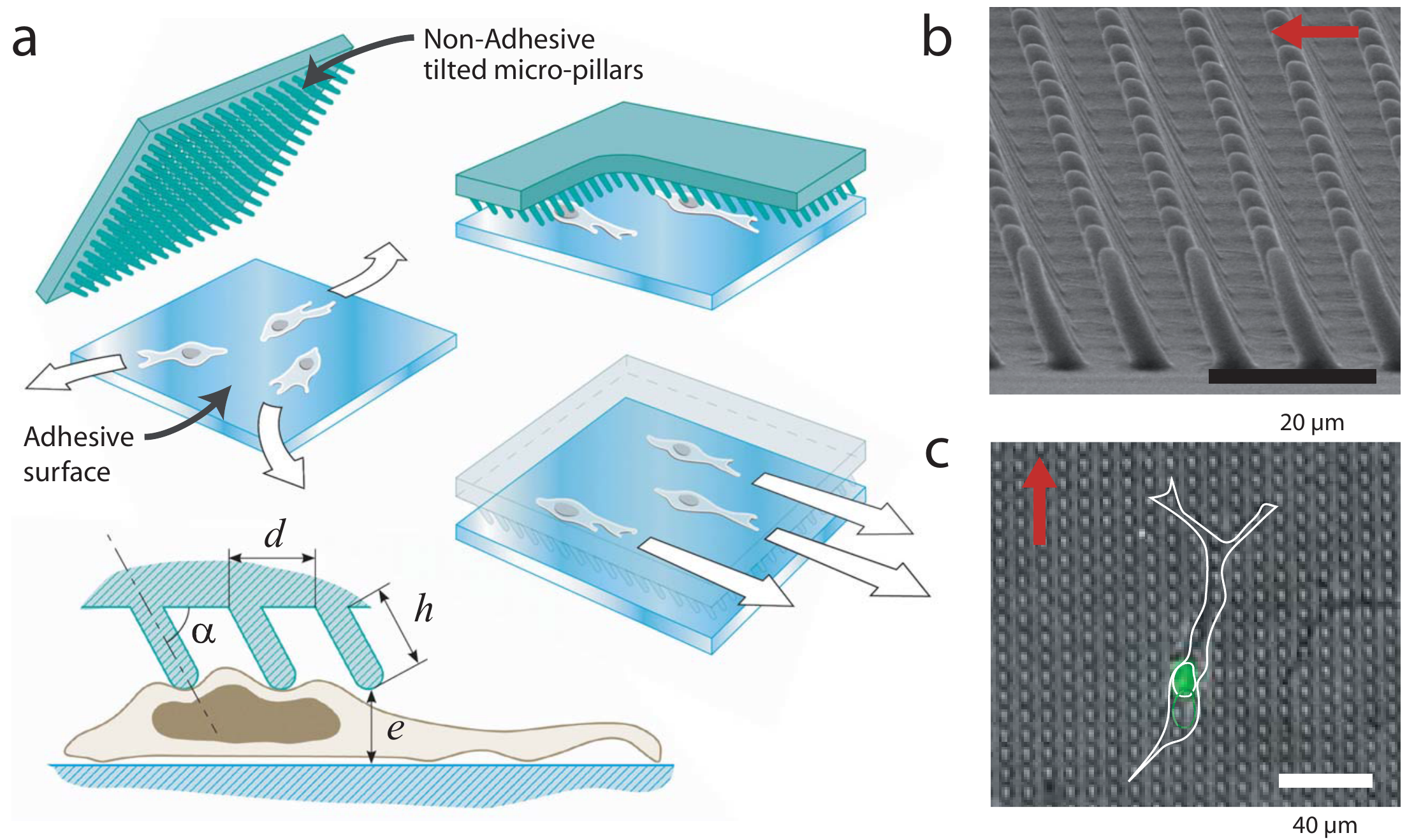}
\caption{Cell migration assay under non-adhesive tilted micro-pillars. a. Design of the assay (see SupFig. 3). b. SEM image of tilted pillars. $\alpha = 70^o,\  d = 10 \mu m,\  h = 12 \mu m$.  c. Overlay of phase contrast image and HOECHST staining (nucleus, green) of a NHDF cell confined under tilted pillars.}
\label{fig1}
\end{figure}
We  designed surfaces covered by tilted pillars, for which it was possible to vary independently different geometrical parameters (pillars length, tilt and spacing, Fig. \ref{fig1}a). We  characterized the migration behavior of Normal Human Dermal Fibroblasts (NHDFs) under these surfaces and analyzed the movements of their nuclei, which revealed a strong bias largely independent of the specific geometry of the substrate. Our analysis reveals that most features of cell trajectories, including the bias,  can be predicted by a simple  model of active Brownian particle in a ratchet potential, which we suggest originates from the elastic interaction of the cell body with the environment.  We stress that this interaction, resulting in an effective  asymmetric  friction of purely geometric nature, is non adhesive,  highly non-specific and directly tunable by varying geometrical parameters. We therefore believe that the guiding effect is robust, and could be used as a versatile tool to direct cell migration both in vitro and in vivo.

We used a soft lithography method \cite{Velve-Casquillas:2010nx} to produce Polydimethylsiloxane (PDMS) pads covered with tilted pillars (Fig. \ref{fig1}a-b, and Supplementary Information (SI)), the tilt being responsible for the anisotropy of the micro-structures. We produced square array of pillars, each $2\ \mu m$ in diameter, with different spacing $d$, lengths $h$, and tilt angles $\alpha$. The pillars, which can be considered as undeformable in experimental conditions,  were made non-adhesive by coating their surface with Polyethylene Glycol (PEG). We plated NHDFs on to a fibronectin-coated PDMS surface with large, $5\ \mu m$ high spacers and then placed the pillar-bearing pad on to the spacers (Fig. \ref{fig1}a and SI). Cells migrated between the two surfaces, adhering on their ventral side to the fibronectin-coated bottom surface and in contact with the non-adhesive pillars on the top (Fig. \ref{fig1}c and SI). Migrating cells, whose density was low enough to be considered as independent (Supplementary Movies S1 and S2)  were recorded for $33$ hours ($H$) and  cell nuclei were automatically tracked.

\begin{figure}
\includegraphics[width=8cm]{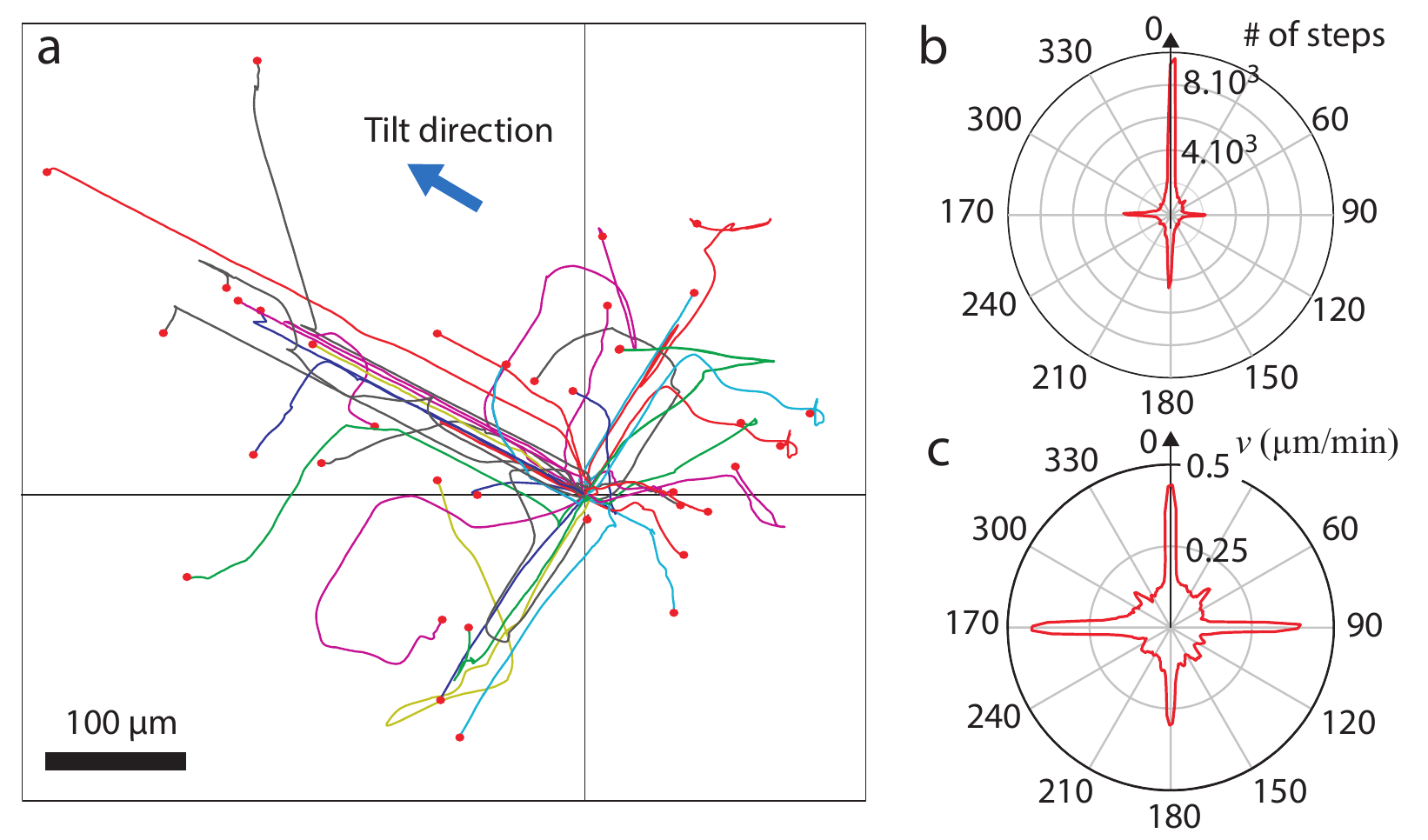}
\caption{ a. Typical tracks (starting at the origin) of cells moving under tilted pillars.  b. Angle histogram of the instantaneous direction of migration (N = 252 cells from 12 movies, n = 78800 single steps).  c.  median speed as a function of the instantaneous direction of migration. (N = 452 cells from 12 movies, n = 100949 single steps). In b,c $0^o$ is the direction of pillar tilt. In a-c pillars geometry is $d = 10 \mu m,\  h = 10 \mu m,\  \alpha = 65^o$}.
\label{fig2}
\end{figure}

\begin{figure}
\includegraphics[width=8.2cm]{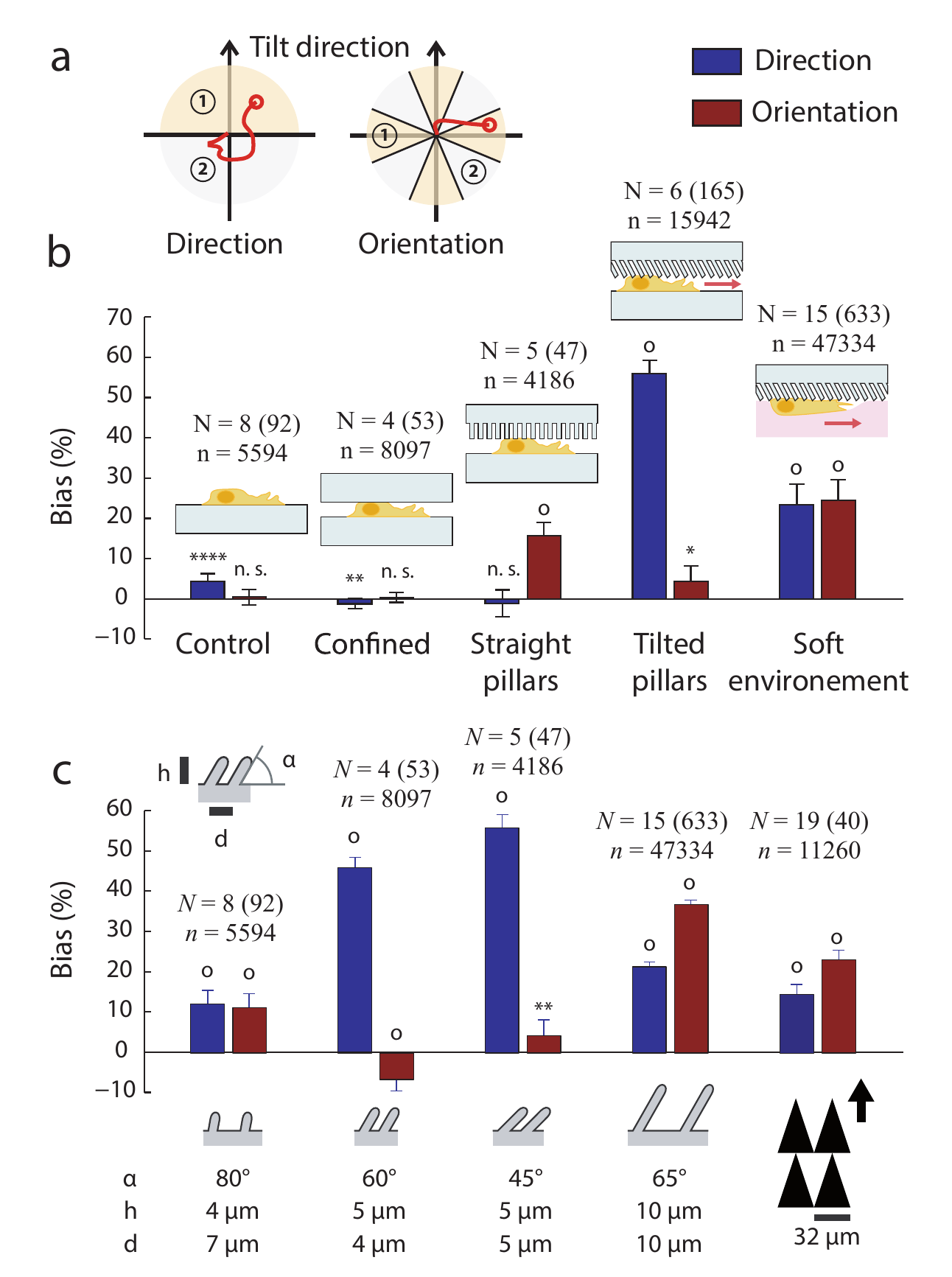}
\caption{ a. Measurement method for the direction and orientation bias quantified in b and c.  Bias is calculated with: Bias = $(\#_1-\#_2)/(\#_1+\#_2)$ where $\#_i$ is the number of cells scored in region $i$ around their initial position after $4H$. 100\% (-100\%) means that all cells are in region 1 (2) after $4H$.   b. Migration bias in controls and for different migration set-ups. Pillar geometries $(d;h;\alpha)$  are: $(5 \mu m; 5 \mu m; 90^o)$ for straight pillars, $(5 \mu m; 5 \mu m; 45^o)$ for tilted pillars and $(5 \mu m;6.5 \mu m;60^o)$ for soft environment. c. Migration bias for cells confined under tilted pillars of various geometries and under micro-prisms of $2 \mu m$ in depth, $3 \mu m$ above cells.  In b,c $N$ corresponds to the number of cells, $n$  to the number of single steps. Significance of the bias are given according to the $\chi^2$ test, on the single steps: * is $p<0.05$,  ** is $p<0.01$, **** is $p<0.0001$, o is $p<10^{-8}$. }
\label{fig3}
\end{figure}

\begin{figure}
\includegraphics[width=9cm]{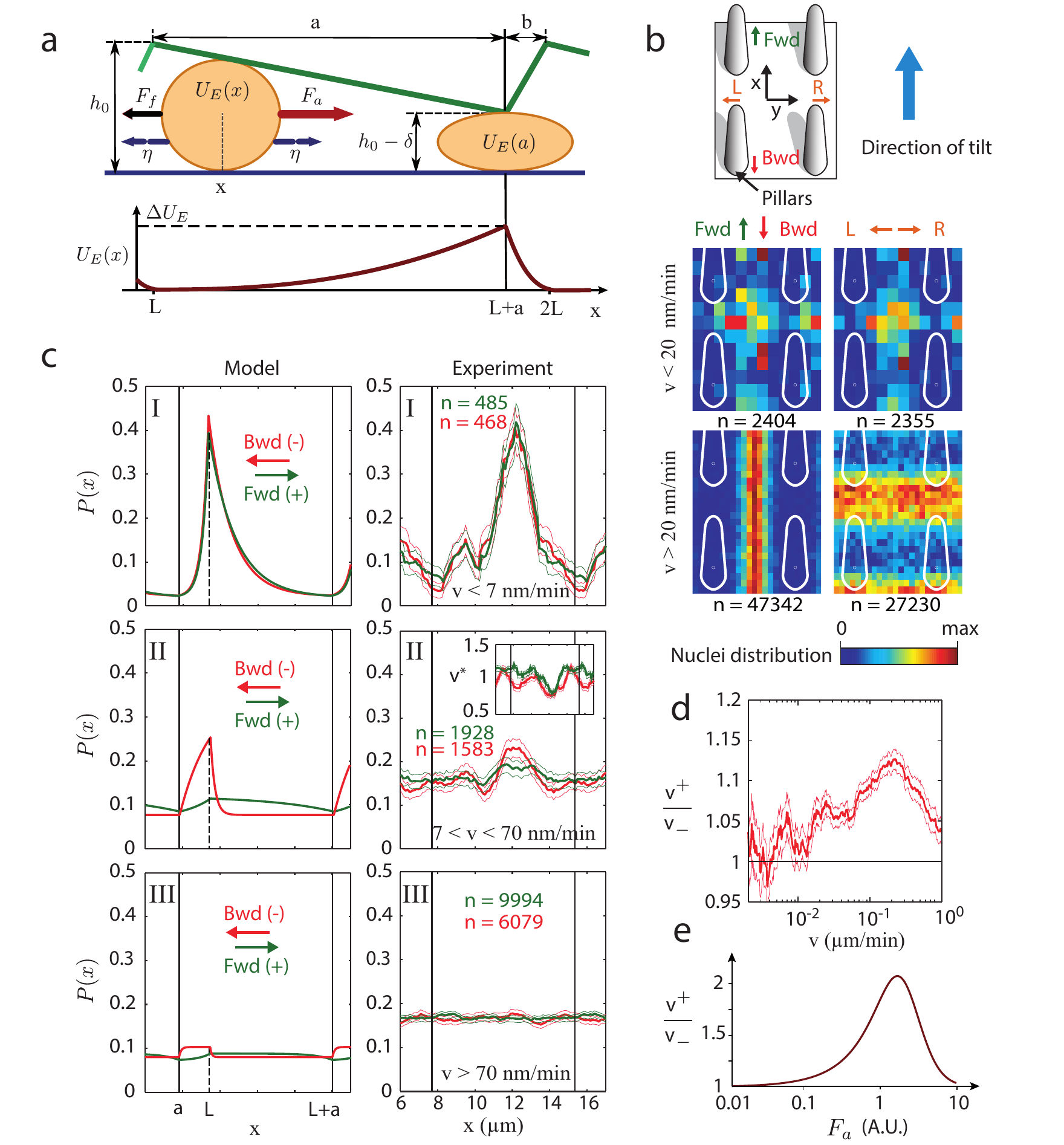}
\caption{ a. Model: deformation of an elastic sphere (the cell nucleus) in a sawtooth  profile (top) and the corresponding energy profile (bottom). b. Top view drawing of pillars and definition of axes and directions  used in  distribution maps (bottom). Fwd, Bwd, L and R give the directions used in the remaining of the figure. Experimental distribution maps of the center of mass of nuclei as a function of the cell speed and axis of migration (bottom). c. Theoretical distribution of nuclei predicted by the model (left) and measured (right) for the 3 speed (force for the model) regimes I-III. Vertical lines indicate the position of the pillars (minimum of $h(x)$, see a.). Inset in the middle right graph shows the median speed of the nucleus as a function of the position of its center of mass and its direction of motion (speeds are normalized by the median speed). Values used in theoretical curves are in regime I:  $FL= 0.1 \Delta U_E$; In II : $FL= 6 \Delta U_E$ ; In III : $FL= 30 \Delta U_E$. d, Speed ration $v_+/v_-$, as a function of  the median instantaneous speed $v$.  e, Model prediction of $v_+/v_-$ as a function of rescaled force $FL/\Delta U_E$. Data set used for all graphs is the same as in Fig. 2.  }
\label{fig4}
\end{figure}

Typical paths (Fig. \ref{fig2}a) showed two clear features: cells tended to follow the two main axes of the lattice (quantified as orientation bias in Fig \ref{fig3}) and they migrated more in the direction of the tilt of the pillars (direction bias in Fig \ref{fig3}), which is the central effect discussed in this paper. To assess the robustness of the guiding effect and the role of specific parameters of pillar geometry, we quantified the orientation and direction bias after $4 H$ of migration, while varying $d$, $h$ and $\alpha$. In control experiments, when cells were not confined, or were confined with a non-adhesive flat surface, their mean square displacement was characteristic of an isotropic persistent random motion (see SI).  
Cells covered with straight pillars showed a strong orientation bias along the lattice two main axes, but no direction bias. When cells were topped by tilted pillars, they showed a significant direction bias in all tested conditions (Fig \ref{fig3}a-c). The direction bias was maximal in the case shown in Fig \ref{fig3}c ($d=5 \ \mu m$, $l=5 \ \mu m$, $\alpha=45^0$), where 78\% of cells migrated more in the direction of the tilt after $4 H$. A detailed analysis of cell migration paths (see Fig  \ref{fig2}a for a sample) revealed that both the fraction of  time spent migrating in a given direction (Fig. \ref{fig2}b) and  the speed  as a function of direction (Fig. \ref{fig2}c)  were strongly biased and contributed to the overall observed direction bias : cells were  slower in the backward (Bwd) direction (defined relative to pillar tilt, see Fig \ref{fig4}b) indicating that they effectively experienced more friction in this direction, and also   spent  more time in the forward (Fwd) direction.
When pillars were spaced by more than $7 \ \mu m$, orientation bias was strong, as cells tended to follow the main axes of the  lattice. This effect appeared when spacing was large enough to allow the cell nuclei, whose diameter is of about $14 \ \mu m$, to partially squeeze between pillars (Supplementary Movie S1). This observation suggests that mechanical interaction of the cell body, which mostly contains the nucleus, with the covering anisotropic non-adhesive features has an important role in the migration behavior. Overall, by systematically varying the geometry of the pillars, we found that the direction bias was a robust feature that increased with pillar density and with tilt (Fig. \ref{fig3}c), while the pillar spacing affected orientation bias.  We focus below on the direction bias.

We also tested anisotropic periodic features with completely different geometries, such as micro-prisms (Fig \ref{fig3}c and  SI), and changed the properties of the bottom substrate on which cells migrated, by applying tilted pillars over cells plated on a soft poly-acrylamide gel ($15 kPa$ stiffness) coated with fibronectin (Fig \ref{fig3}b). Such soft substrates are closer to conditions which might be experienced by migrating cells in tissues. In all tested cases, we observed a strong direction bias, which was correlated with the main anisotropic axis of the micro-features, showing the robustness of the effect. 

To investigate the mechanical origin of this biased cell migration, we developed a minimal theoretical model of cell motion in a ratchet profile geometry. The model assumes that  cell motion is restricted along the $x$ axis on a flat surface and is confined by a rigid top of periodic asymmetric profile $h(x)$ (Fig. \ref{fig4}a). For modeling purposes, we  considered a piecewise linear shape of maximum height $h_0$ and minimal height $h_0-\delta$ ; results being largely independent of this choice. The period is denoted by $L = a + b$, and the profile $h(x)$ assumed to be successively decreasing  over a distance $a$ and then increasing  over a distance $b$. Based on the observations presented above, it is  hypothesized that cell movement is mainly restricted by the interaction between the geometric features of the top surface and the cell body, which has to deform to pass through the successive bottlenecks. The cell body, composed mainly of the nucleus is assumed to be a spherical linear elastomer \cite{Pajerowski:2007oq} of Young modulus $E$ and equilibrium radius $R$. It can be shown (see SI) that the elastic deformation imposed by the geometry when  $h(x)<2R$ then leads to an elastic energy \cite{landaumeca} stored in the cell body that reads $U_E(x)\propto E\sqrt{R}(2R-h(x))^{5/2}$, and whose maximal variation, denoted $\Delta U_E$, is reached at each bottleneck. It is next assumed that the cell  exerts an active propulsion force $F_a$ on its own cell body, which reads $F_a = \epsilon F + \eta(t)$ where $F$ is a positive constant. Here $\epsilon$ denotes the direction of motion and takes values in $\{-1,+1\}$ (+ and -  corresponding here to the Fwd and Bwd directions defined above); it reflects the cell polarity and is assumed to randomly leave state $\pm 1$ with rate $\nu_\pm$. The white noise $\eta(t)$  reflects the stochasticity of cell  protrusion dynamics, which is assumed to be much faster that  the typical polarization  time $1/\nu_\pm$.    
The dynamics of the position $x(t)$ of the cell body is then given in the over-damped regime by a classical Langevin equation
\begin{equation}
\lambda d_t x=\epsilon F-\partial_x U_E+\eta=-\partial_x U_E^\epsilon+\eta
 \end{equation}
where  $U_E^\epsilon \equiv U_E-\epsilon Fx$, $\lambda$  (assumed constant) effectively accounts for the friction (assumed linear) of the cell body with its environment  and $\epsilon$ follows the random telegraph dynamics prescribed above ; this can be seen as an example of active Brownian motion \cite{Romanczuk:2011}, which already proved to be successful to model cell trajectories \cite{Selmeczi:2005uq,Selmeczi:2008vn}. Following the observation that  persistence lengths are much larger than a lattice step, it can be assumed that the cell is transiently at steady state after each direction change. For each direction $\epsilon$ fixed, the problem is then formally equivalent to a Brownian particle in a tilted potential $U^\epsilon_E$ (or tilted washboard), which has been studied in various contexts in physics\cite{Reimann:2002ve}. The mean velocity  $v_\epsilon$ and steady state probability distribution of the cell body $P_\epsilon (x)$ in the landscape  can then be calculated exactly \cite{H.Risken:1996} (see SI for explicit expressions). Finally   the bias can be quantified in the long time limit by
\begin{equation}\label{bias}
\frac{\langle x\rangle}{t}\sim\frac{\nu_- v_+ -\nu_+ v_-}{\nu_++\nu_-}.
\end{equation}
Let us assume that $\nu_+=\nu_-$. Motion in a flat potential ($U_E=0$) is then obviously symmetric ($\langle x\rangle=0$). Interestingly, Eq.(\ref{bias}) shows that motion can nevertheless be biased as soon as $v_+\not=v_-$, which is realized for asymmetric potentials ($a\not=b$) as we proceed to show, and is in agreement with \cite{Angelani:2011fk}.

Depending on the height of the energetic barrier $\Delta U_E$ imposed by the geometry as compared to the work $FL$ performed by the active force over a period, three different regimes (denoted hereafter I-III) can be predicted. For strong geometric constraints $\Delta U_E\gg FL$, we find that the cell body is blocked by energy barriers of comparable height in both + and - directions (for small enough noise intensity), resulting in a negligible velocity in both directions (regime I) and therefore a negligible bias.  The expected distribution  $P_\epsilon(x)$  is then markedly peaked at the position of the minimal energetic constraint ($x=0$) and almost independent of the direction (Fig. \ref{fig4}c top left). The key point is that for intermediate geometric constraints $\Delta U_E\sim FL$  (regime II), the model shows that for  $a>b$  the cell has to overcome a barrier that is significantly lower in the + direction ($\Delta U_T^+=\Delta U_E-Fa$) than in the - direction ( $\Delta U_T^-=\Delta U_E-Fb$). This results in a clear anisotropy of the mean velocity quantified analytically by the ratio $v_+/v_-$ (see Fig. \ref{fig4}e and SI for explicit expressions), which we call here geometric friction. A significant bias is therefore obtained in this regime, as shown by Eq. (\ref{bias}). The expected distribution $P_\epsilon(x)$  is then peaked at the position of the minimal energetic constraints, with however a significantly smaller peak in the preferred direction + (Fig. \ref{fig4}c middle left). Last, if geometric constraints are too weak $\Delta U_E\ll FL$  (regime III), energetic barriers become negligible, isotropy is recovered and no bias is observed. In this regime, $P_\epsilon(x)$  is almost uniform in both directions. 

To assess the relevance of this model to our experiments, we performed extensive statistical analyses of the positions of cell nuclei, relative to the surrounding micro-pillars, along migration paths (Fig. \ref{fig4}b-d). Sets of four micro-pillars were systematically registered and the positions of the centers of mass of nuclei relative to the closest surrounding pillars were computed for each time point and each cell (density probability, $P(x,y)$, is shown in Fig \ref{fig4}b). Positions of nuclei were then sorted according to instantaneous cell speed and instantaneous cell direction (Fig. \ref{fig4}b). 
The analysis of  the variation in nuclei density  along the $x$-axis as a function of the median cell speed (which  yields an indirect measure of the active force) confirmed the three regimes predicted by the model, and the theoretical and experimental distribution profiles were in good qualitative agreement in each case  (Fig. \ref{fig4}c). In particular, for intermediary speeds ($7\  nm/min < v < 70\  nm/min$, regime II), our experiments showed a clear difference between the Fwd  and  Bwd directions (Fig \ref{fig4}c middle panels). Furthermore, in this regime II, the maxima of $P(x)$ (the density probability of nuclei positions averaged along the y axis) clearly corresponded with a decrease in speed (Fig. \ref{fig4}c, insert in middle right panel), suggesting that the micro-structures are blocking the nuclei. 
 Last, the experimental measure of the ratio between Fwd and Bwd speed ($v_+/v_-$), as a function of median cell speed (Fig. \ref{fig4}d) showed the same non monotonic profile predicted by the model (Fig. \ref{fig4}e). Similar results were obtained in three independent experiments using tilted pillars, but also using micro-prisms (SI). All together, these analyses strongly indicate that the cell body behaves as an active particle in a   ratchet potential that originates from its mechanical interaction with the environment, and  which we suggest is the mechanism responsible for the observed bias. This observation could be related to \cite{Guo:2011kl}, which showed that cells injected through microscale funnel constrictions  experienced an asymmetric energy profile . 
 
In conclusion, the theoretical analysis enables the characterization of the regime (II) in which  a strong bias is observed. 
The  model  offers an accurate prediction of the difference in speed between the Fwd and the Bwd directions and therefore the bias, even for symmetric direction changes $\nu_+=\nu_-$. The longer persistence time observed experimentally in the Fwd direction, which could be related to the increased migration speed since these two parameters of cell migration are correlated \cite{Maiuri:2012tg}, suggests that $\nu_\pm$  could for example depend on speed.  Overall, our work strongly supports the concept of geometric friction as a guiding force for migrating cells constrained by non-adhesive substrates: cells prefer to move, and move faster, in the "smoothest" direction.  Geometric friction is shown to rely on simple and non-specific ingredients such as an anisotropic landscape and an elastically deformable cell body, and is therefore expected to be robust.  This robustness   suggests that geometric friction might be a general way by which migrating cells could sense anisotropy in their environment. It could be used as a versatile tool to direct cell migration both in vitro and in vivo, or  investigate the role of mechanical properties of the nucleus in cell migration.

%\bibliography{../../../biblio/liste_live}
%\bibliographystyle{../../../biblio/vincent}
%\newpage

%\textit{thanks}
\begin{acknowledgments}
We acknowledge the PICT-IBISA microscopy platform at Institut Curie. We also acknowledge D. Riveline, C. Baroud et J.F. Joanny for discussion and comments on the manuscript. Vivatech and ARC are acknowledged for their financial support to M LB and YJ L and ANR-09-PIRI-0027 to MP and RV.
\end{acknowledgments}

\end{document}